\documentclass[12pt]{article}
\usepackage{graphicx}
\begin{document}
\rightline{NKU-2012-SF3}
\bigskip
\begin{center}
{\Large\bf Quasinormal Modes  of  Bardeen  Black Hole: Scalar Perturbations}

\end{center}
\hspace{0.4cm}
\begin{center}
Sharmanthie Fernando \footnote{fernando@nku.edu} \& Juan Correa \footnote{correaj1@mymail.nku.edu}\\
{\small\it Department of Physics \& Geology}\\
{\small\it Northern Kentucky University}\\
{\small\it Highland Heights}\\
{\small\it Kentucky 41099}\\
{\small\it U.S.A.}\\

\end{center}

\begin{center}
{\bf Abstract}
\end{center}

\hspace{0.7cm} 

The purpose of this paper is to study quasinormal modes (QNM)  of the Bardeen black hole due to scalar perturbations. We have done a thorough analysis of the QNM frequencies by varying the charge $q$, mass $M$ and the spherical harmonic index $l$. The unstable null geodesics are  used to compute the QNM's in the eikonal limit. Furthermore, massive scalar field modes are also studied by varying the mass of the field. Comparisons are done with the QNM frequencies of the Reissner-Nordstrom black hole.


\section{Introduction}

In general, black hole space-times are expected to have horizons as well as singularities covered by the horizons. Contrary to this notion, a ``regular'' 
space-time without a singularity and with a horizon was proposed by Bardeen \cite{bardeen}. This particular paper is not readily available. However, a discussion of this model is given by Borde in \cite{borde1} \cite{borde2}. There were other regular black holes other than the one proposed by Bardeen and they  were all referred to as `` Bardeen black holes'' by Borde \cite{borde1}. In this paper, we will focus on the space-time proposed by Bardeen \cite{bardeen}.

Ay\'{o}n-Beato and Garc\'{i}a \cite{gar4} proposed a model of nonlinear electrodynamics coupled to Einstein gravity to obtain Bardeen black hole as an exact solution. Hence, the Bardeen black hole can be interpreted as the solution to a nonlinear magnetic monopole with a mass $M$ and a charge $q$ \cite{gar4}.
Ay\'{o}n-Beato and Garc\'{i}a \cite{gar4} have presented several other interesting regular solutions with  nonlinear electrodynamics coupled to General Relativity in \cite{gar2}\cite{gar3}\cite{gar4}\cite{gar5}.

There are several works in the literature related to the Bardeen black hole.
Gravitational lensing of the regular black hole was studied by Eiroa and Sendra \cite{eiroa1}. The geodesic structure of the test particles around the Bardeen black hole were studied by Zhou et.al \cite{zhou}. Gravitational and electromagnetic stability were discussed by Moreno and Sarbach \cite{olivier}. Quantum corrections for the Bardeen black hole was presented by Sharif and Javed \cite{javed}.

In this paper, our focus is on studying the scalar field perturbations of the Bardeen black hole and to compute the quasinormal modes of the perturbations.

When a black hole undergo perturbations, the resulting behavior can be described in three stages. The first stage corresponds to  radiation due to the initial conditions of the perturbations. The second stage corresponds to  damped oscillations with complex frequencies. These frequencies are independent of the initial conditions and are only dependent on the black hole properties such as the mass, charge and the angular momentum. These modes are called quasinormal modes (QNM). The third stage in general corresponds to a power law decay of the fields. 

QNM's have attracted lot of attention from the research community. First of all, there is interest  from the experimental point of view, since, there is hope that the QNM's  may be detected  by the gravitational antennas such as LIGO, VIRGO and LISA in the future. Since the QNM's only depend on the properties of the black holes, such detections would give clues to identify the physical properties of the black holes. On the other hand, due to the famous relation between AdS/CFT duality, many works have focused on studying QNM's of black holes with a negative cosmological constant \cite{mann} \cite{horowitz}. Another reason to create interest on QNM's was the conjecture by Hod \cite{hod3} relating quantum properties of Schwarzschild black hole and asymptotic QNM's. There are many works aimed at computing asymptotic QNM frequencies along those lines. An excellent review on QNM's is written by Konoplya and Zhidenko \cite{kono1}. It is fascinating to see some new work relating QNM to various aspects of black hole physics.  For example, some recent work have addressed relation between QNM's and hidden conformal symmetry \cite{chen} \cite{kim1}. Another interesting paper was written on the connection between gravitational lensing and QNM's by Stefanov et.al \cite{stef}. Not only black holes, even the naked singularities have been studied from the QNM point of view \cite{saa}. Given all the  above, it is worthwhile to seek answers how nonlinear sources modify the QNM properties of a black hole.

The paper is presented as follows: In section 2, the Bardeen black hole solutions are introduced. In section 3, the  perturbations by a massless scalar filed is given. In section 4, we will computer the QNM's using the sixth order WKB approach and discuss the results. In section 5, a relation between the null geodesics and the QNM's are presented. In section 6, the perturbations by a massive scalar field is studied. The summary is given in section 7. Directions for further studies are given in section 8.

\section{Introduction to the regular Bardeen black hole}

In this section, we will give an  introduction to the regular static charged  black hole named as Bardeen black hole \cite{bardeen}. Ay\'{o}n-Beato and Garc\'{i}a \cite{gar1} interpreted the Bardeen black hole as the gravitational field of a magnetic monopole arising from non-linear electrodynamics. The proposed action to include the non-linear electrodynamic term is,
\begin{equation}
S = \int d^4x \sqrt{-g} \left[ \frac{R }{16 \pi G} - \frac{ 1}{ 4 \pi} {\cal L}(F) \right]
\end{equation}
Here, $R$ is the scalar curvature, and ${\cal L}(F)$ is a function of $F = \frac{1}{4}F_{\mu \nu} F^{\mu \nu}$. Here, $ F_{\mu \nu} = 2 ( \bigtriangledown_{\mu} A_{ \nu} - \bigtriangledown_{\nu} A_{ \mu}) $ is the electromagnetic field strength. In \cite{gar1} the authors derived the function 
$ {\cal L}(F)$ in order to obtain the Bardeen black hole as,
\begin{equation}
{\cal L} (F) = \frac{ 3}{ 2 s q^2} \left( \frac{ \sqrt{2 q^2 F}}{ 1 + \sqrt{ 2 q^2 F}} \right)^{\frac{5}{2}}
\end{equation}
Here, $q$ and $M$ are the magnetic charge and the mass of the magnetic monopole. Also,
$s = \frac{ |q|}{ 2 M}$. The equations of motion derived from the action in eq.(1) is given by,

\begin{equation}
G_{\mu}^{\nu} = 2 \left( \frac{ \partial {\cal L}}{ \partial F} F_{\mu \lambda} F^{ \nu \lambda} - \delta_{\mu}^{\nu} {\cal L} \right)
\end{equation}

\begin{equation}
\bigtriangledown_{\mu} \left(\frac{ \partial {\cal L}}{ \partial F}  F^{\nu \mu}\right)=0
\end{equation}
Static spherically symmetric solution for the above equations were proved to be the Bardeen black hole solution given by the metric,

\begin{equation}
ds^2 = -f(r) dt^2 + f(r)^{-1} dr^2 + r^2 ( d \theta^2 + sin^2(\theta) d \varphi^2)
\end{equation}
where,
\begin{equation}
f(r) = 1 - \frac{2M r^2}{ ( r^2 + q^2 ) ^{3/2} }
\end{equation}
The magnetic field is given by,
\begin{equation}
F_{\theta \varphi } =  2 q sin \theta
\end{equation}
For $q \neq 0$, the space-time in eq.(5) has horizons only if, 
$ |q| \leq \frac{ 4 M}{ 3 \sqrt{3}}$. This was shown by Borde \cite{borde1} \cite{borde2}. For $ q > \frac{ 4 M}{ 3 \sqrt{3}}$, there are no horizons. For $ q = \frac{ 4 M}{ 3 \sqrt{3}}$, there are degenerate horizons. The function $f(r)$ is plotted in Fig.1 for varying magnetic charge $q$.

\begin{center}
\scalebox{.9}{\includegraphics{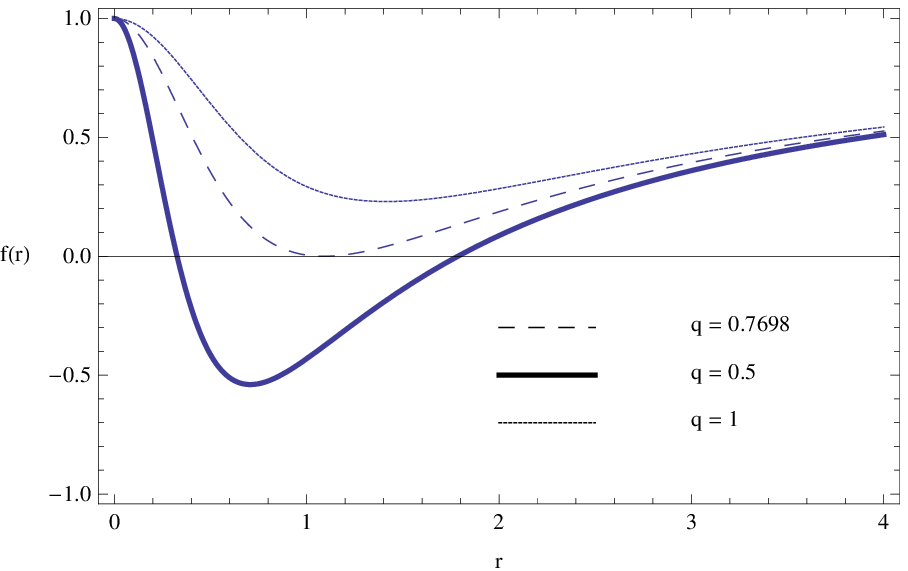}}

\vspace{0.3cm}
\end{center}

Figure 1. The figure shows the function $f(r)$  for $M=1$ and varying values of $q$. \\

Asymptotically, the metric function $f(r)$ behaves as,
\begin{equation}
f(r) \approx 1 - \frac{ 2 M}{r}  + \frac{3 M q^2}{ r^3} + O \left( \frac{ 1}{ r^5} \right)
\end{equation}

The metric for the black hole  in Einstein-Maxwell gravity, given by the well known Reissner-Nordstrom black hole, with a magnetic charge is,

\begin{equation}
ds^2 = -f(r)_{RN} dt^2 + f(r)_{RN}^{-1} dr^2 + r^2 ( d \theta^2 + sin^2(\theta) d \varphi^2)
\end{equation}

where,

\begin{equation}
f(r)_{RN} = 1 - \frac{2 M}{r} + \frac{ q^2}{r^2}
\end{equation}
In Fig.2, the two metric functions for the Bardeen black hole and the Reissner-Nordstrom black hole are plotted for comparison. It is clear from the Fig.2, that both black holes have two horizons. For small $r$, the behavior is some what different even though asymptotically, both functions are similar. The non-singular nature of the function $f(r)$ for the Bardeen black hole is observed from the Fig.2. As discussed in  \cite{gar1}, the space-time is regular everywhere since all the scalar curvatures, $ R$, $ R_{\mu \nu} R^{ \mu \nu}$ and $ R_{\mu \nu \alpha \beta} R^{ \mu \nu \alpha \beta }$ are regular every where. However, the electromagnetic invariant $ F = \frac{g^2}{ 2 r^4}$ has singular behavior. The Bardeen black hole satisfy the weak energy condition.

\begin{center}
\scalebox{.9}{\includegraphics{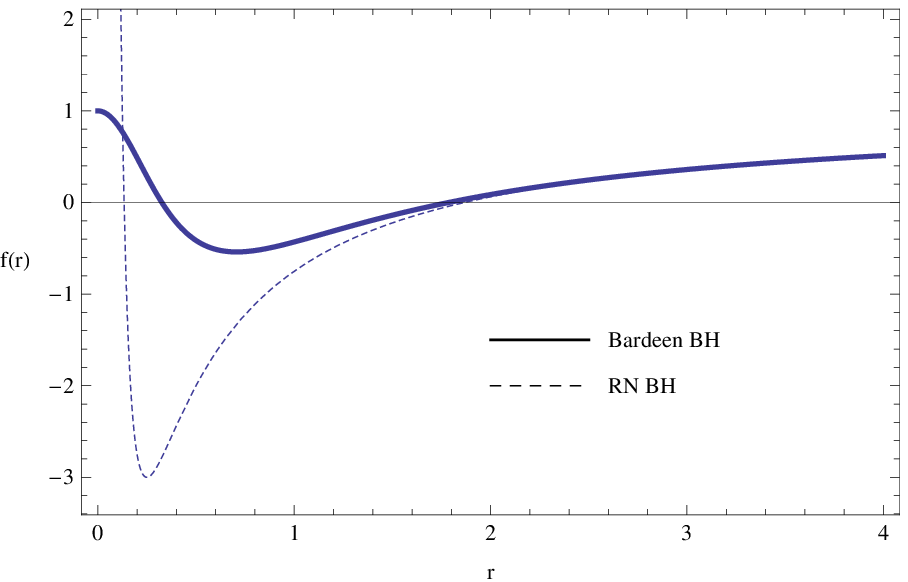}}

\vspace{0.3cm}
\end{center}

Figure 2. The figure shows the function $f(r)$  for the Bardeen black hole(dark) and the Reissner-Nordstrom black hole(dashed). Here, $M=1$ and $q = 0.5$.\\

The Hawking temperature of the Bardeen black hole is given by,
\begin{equation}
T = - \frac{1}{4\pi} \left. \frac{ d g_{tt}}{d r} \right|_{r = r+} = \frac{ 1 }{ 4 \pi} \left[ \frac{ 2 Mr_{+} ( r_{+}^2 - 2 q^2)}{ ( q^2 + r_{+}^2)^{5/2} ) } \right]
\end{equation}
Here, $r_+$ is the event horizon of the black hole which is a solution of $f(r)=0$. In comparison, the Hawking temperature for the Reissner-Nordstrom black hole is,
\begin{equation}
T_{RN} = \frac{ 1}{4 \pi} \left[ \frac{ 2 M}{ r_{+}^2} - \frac{ 2 q^2}{ r_{+}^3} \right]
\end{equation}
The temperature for both black holes for the same mass is plotted in Fig.3. The Reissner-Nordstrom black hole is ``hotter" than the Bardeen black hole.
\begin{center}
\scalebox{.9}{\includegraphics{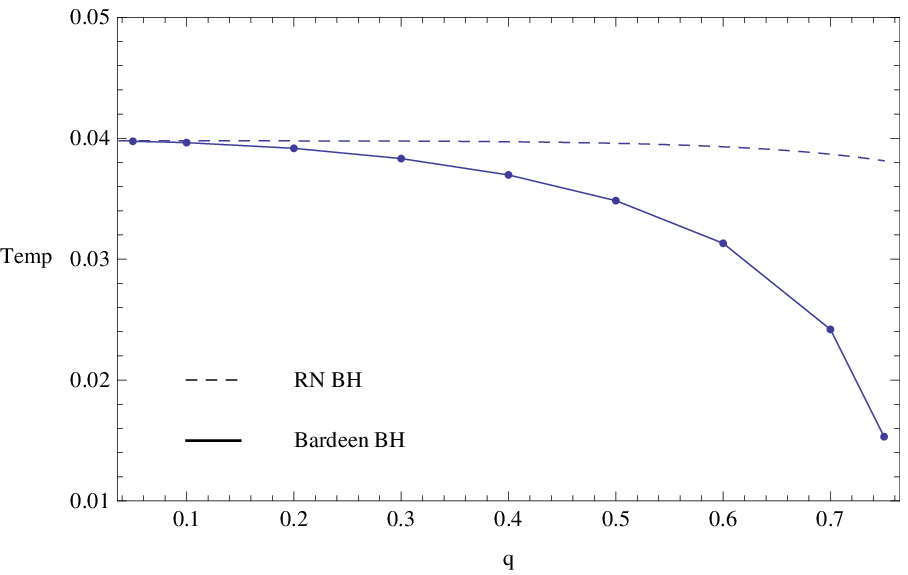}}

\vspace{0.3cm}
\end{center}

Figure 3. The figure shows the temperature $T$  for the Bardeen black hole(dark) and the Reissner-Nordstrom black hole(dashed) as a function of the magnetic charge $q$. Here, $M=1$.

\section{Masslesss scalar perturbation of Bardeen black holes}

In this section, we will introduce scalar perturbation by a massless field around the Bardeen black hole. The Klein-Gordon  equation for a massless scalar field $\Phi$ in curved space-time can be written as,
\begin{equation}
\bigtriangledown ^2 \Phi  =0
\end{equation}
which is equal to,
\begin{equation}
\frac{1}{\sqrt{-g}} \partial_{\mu} ( \sqrt{-g} g^{\mu \nu} \partial_{\nu} \Phi ) =0
\end{equation}
Using the ansatz for the scalar field $\Phi$,
\begin{equation}
\Phi =  e^{- i \omega t} Y(\theta,\phi) \frac{\xi(r)}{r} 
\end{equation}
eq.(14) simplifies to the Schr\"{o}dinger-type equation given by,
\begin{equation}
\frac{d^2 \xi(r)}{dr_{*}^2} + \left( \omega^2  -  V(r_*) \right) \xi(r) =0
\end{equation}
where,
\begin{equation}
V(r) =  \frac{l(l+1) f(r)}{r^2}  + \frac{f(r) f'(r) }{r} 
\end{equation}
Here, $r_*$ is the  well known ``tortoise'' coordinate given by,
\begin{equation}
dr_{*} = \frac{dr}{f(r)}
\end{equation}
Note that $l$ is the spherical harmonic index. 
Here, $r_*$ cannot be evaluated explicitly due to the nature fo the function $f(r)$.  When $r \rightarrow \infty$, $r_* \rightarrow \infty$ and when $r \rightarrow r_+$, $r_* \rightarrow - \infty$.

The effective potential $V(r)$ for the Bardeen  black hole is plotted to display how it changes with charge $q$,  the mass $M$, and the spherical harmonic index $l$  in Fig.4, Fig.5 and Fig.6 respectively.

\newpage

\begin{center}
\scalebox{.9}{\includegraphics{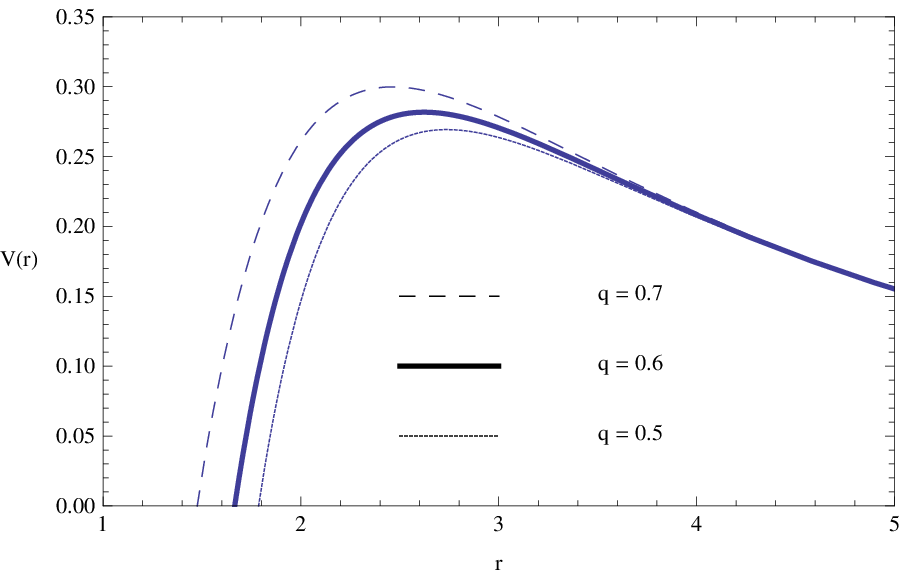}}

\vspace{0.3cm}
\end{center}

Figure 4. The behavior of the effective potential $V(r)$ with 
the charge for the Bardeen black hole. Here, $M=1$ and $l=2$. The height of the potential decreases when the charge decreases.

\begin{center}
\scalebox{.9}{\includegraphics{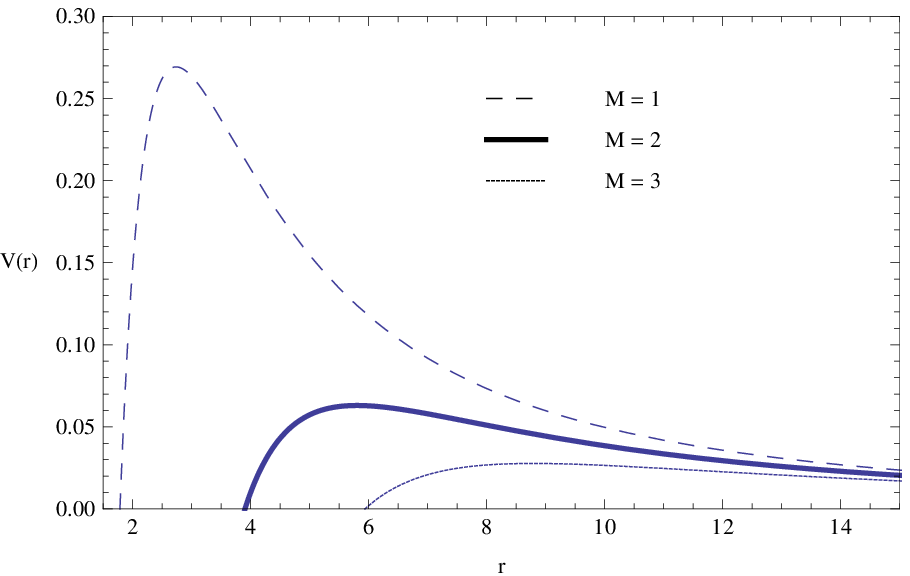}}

\vspace{0.3cm}
\end{center}

Figure 5. The behavior of the effective potential $V(r)$ with 
the mass $M$. Here, $q=0.5$ and $l=2$. The maximum height of the potential increases as $M$ decreases.

\newpage

\begin{center}
\scalebox{.9}{\includegraphics{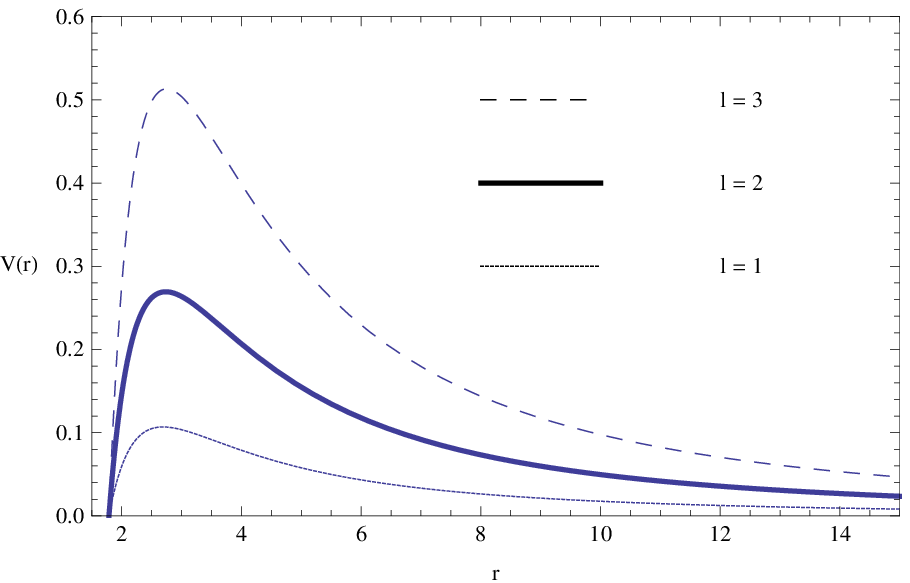}}

\vspace{0.3cm}
\end{center}

Figure 6. The behavior of the effective potential $V(r)$ with 
the spherical harmonic index $l$. Here, $M=1$ and $q = 0.5$. The height of the potential increases when $l$ increases.

We have also plotted the scalar effective potential for the Reissner-Nordstrom black hole with the one for the Bardeen black hole in Fig. 7 for comparison.

\begin{center}
\scalebox{.9}{\includegraphics{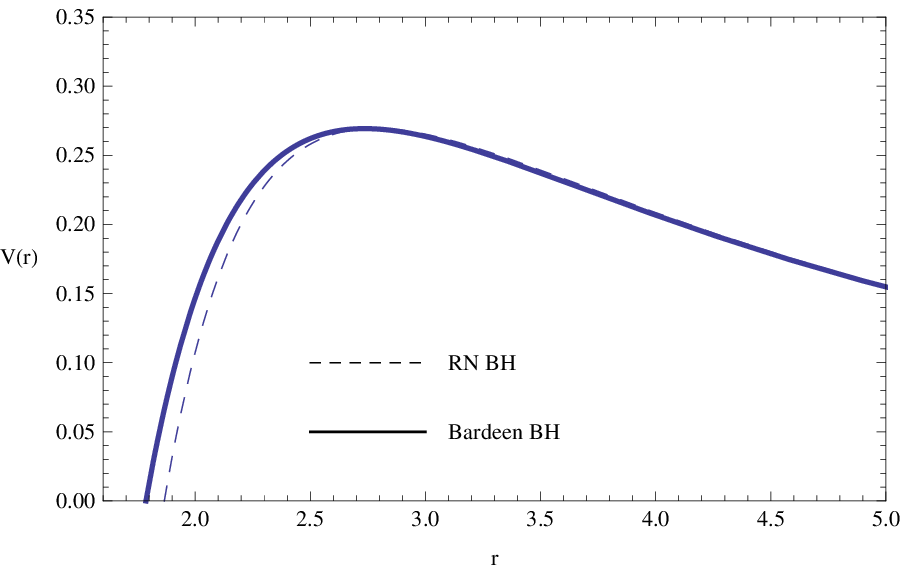}}

\vspace{0.3cm}
\end{center}

Figure 7. The behavior of the effective potential $V(r)$ for the Bardeen black hole (dark) in comparison with the potential for the Reissner-Nordstrom black hole(dashed). Here, $M=1$, $q = 0.2$ and $l=2$. The potential for the Bardeen black hole is slightly higher in the small $r$ range.

\subsection{Remarks on the stability of the black hole}

The potentials are real and positive outside the event horizon for all the figures. Hence, following the arguments by Chandrasekhar \cite{chandra} the Bardeen  black holes can be considered stable classically under perturbations by a massless scalar field.

\section{Quasi-normal modes of Bardeen black hole}

Quasi-normal modes (QNM) for a perturbed   black hole space-times are the solutions to the wave equation given in eq.(16). In order to obtain solutions, one has to impose boundary conditions. At the horizons, the boundary condition is such that the wave has to be purely ingoing. In asymptotically flat space-times, such as the Bardeen space-time, the second boundary condition is for the solution to be purely outgoing  at spatial infinity.

Usually, the wave equation for  black hole perturbations  cannot be solved exactly. There are  few cases of exactly solved models known to the authors which are mentioned here. In 2+1 dimensions, the wave equations of  the well known BTZ black hole \cite{bir1}, the charged dilaton black hole \cite{fer3} \cite{fer2}, the Lifshitz black hole \cite{moon} and the Godel black hole \cite{li} has been solved to obtain exact QNM values. In two dimensions, an asymptotically anti-de-Sitter black hole  has been solved exactly \cite{lopez}. In five dimensions,  Nunez and Starinets have obtained exact values for vector perturbations \cite{nun}.

There are many methods developed to compute QNM's in the literature. In this paper, a semi analytical technique developed by Iyer and Will \cite{will} is followed. The method  is based on  the WKB approximation. Iyer and Will developed it  up to third order and later, Konoplya developed it up to sixth order \cite{kono4}. Konoplya computed QNM frequencies of $D$ dimensional Schwarzschild black holes in that paper which also includes a comparison WKB method with varying orders. Examples of the application of the third order WKB to compute QNM' are given in \cite{fer4} \cite{fer5} and of the sixth order WKB is given in \cite{fer6}.

We will follow the formalism presented in the paper by Konoplya \cite{kono4}. In the WKB formula, the QNM frequencies are related to the effective potential in eq.(17) as,
\begin{equation}
\frac{ \omega^2 - V_0}{ \sqrt{ - 2 V_0^{''}}} - L_2 - L_3 - L_4 - L_5 - L_6 = n + \frac{1}{2}
\end{equation}
Here, $V_0$ and $V_0^{''}$ are the maximum potential and the second derivative of the potential where the maximum occurs. The expressions for $L_2$ and $L_3$ are given in \cite{will} and $L_3$, $L_4$, $L_5$ and $L_6$ are given in \cite{kono4}. Here, $n$ is the overtone number. The computed $\omega$ values are complex and are  as  $\omega = \omega_R - i \omega_I$.

First, we have computed QNM frequencies by varying  the charge of the black hole. We have also computed the QNM's for the Reissner-Nordstrom black hole with the same mass and the charge in order to compare. Here, $ n=0$.

\begin{center}
\begin{tabular}{|l|l|l|l|l|r} \hline \hline

 q & $\omega_R$(RN BH)  &  $\omega_I$ (RN BH) & $\omega_R$(Bardeen BH) 
&  $\omega_I$ (Bardeen BH) \\ \hline

0.1 & 0.484455 & 0.0968185  & 0.484470  & 0.0966541 \\ \hline

0.2 & 0.486929 & 0.0969738 & 0.486999 & 0.0963019 \\ \hline

0.3 &  0.491179 & 0.0972258 & 0.491380 & 0.0956563 \\ \hline

0.4 &  0.497411 & 0.0975605 & 0.497895 & 0.0946064 \\ \hline

0.5  & 0.505966 & 0.0979492 & 0.507037 & 0.0929337 \\ \hline

0.6  & 0.517386 & 0.0983318 & 0.519668 & 0.0901727 \\ \hline

0.7  & 0.532561  & 0.0985743 & 0.537388 & 0.0851340  \\ \hline

0.76  & 0.544071  & 0.0985311  & 0.551623 & 0.0796005  \\ \hline

0.8  & 0.553052  & 0.0983443  & ** & **  \\ \hline

0.85  & 0.566148  & 0.0977987  & ** &  **  \\ \hline

\end{tabular}
\end{center}

\vspace{0.3cm}
\begin{center}
\scalebox{.9}{\includegraphics{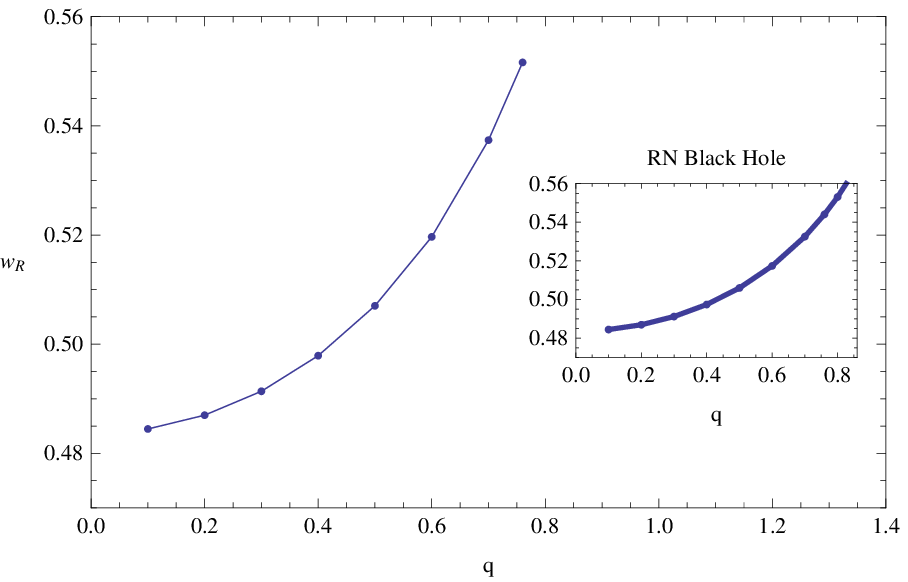}}

\vspace{0.3cm}

\end{center}

Figure 8. The behavior of Re $\omega$ with the magnetic charge  $q$ for $M=1$, and $l=2$.\\

\begin{center}
\scalebox{.9}{\includegraphics{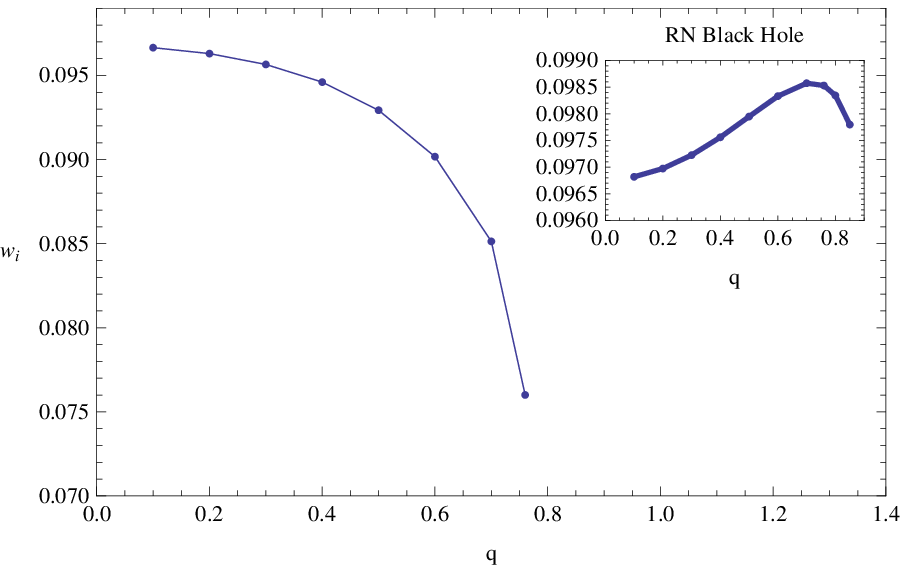}}

\vspace{0.3cm}

\end{center}

Figure 9. The behavior of Im $\omega$ with the magnetic charge  $q$ for $M=1$, and $l=2$.\\

From the above Fig.8 and Fig.9, it is clear that real value of the QNM frequency $\omega$ increases when $q$ increases for both black holes. However, the imaginary part of $\omega$ decreases for Bardeen black hole with charge, while for the Reissner-Nordstrom black hole, there is a maximum before it starts to decrease. We like to mention here that QNM frequencies for Reissner-Nordstrom black hole were studied by Anderson \cite{leaver} and Leaver \cite{ander}. There is a discussion on some of the earlier work on this subject in the book by Frolov and Novikov \cite{fro}.

Next, the QNM values are computed for the Bardeen black hole for various values of the mass $M$ as shown in the following figures, Fig.10 and Fig. 11. Here, $ n=0$.

\begin{center}
\begin{tabular}{|l|l|l|r} \hline \hline
 M & $\omega_R$  &  $\omega_I$ \\ \hline

1 & 0.5070370 & 0.0929337 \\ \hline
2 & 0.2444720 & 0.0480105 \\ \hline
3 & 0.1619870 & 0.0321495 \\ \hline
4 & 0.1212350 & 0.0241475 \\ \hline
5 & 0.0968939 & 0.0193308 \\ \hline
6 & 0.0807027 & 0.0161148 \\ \hline
7 & 0.0691519 & 0.0138156 \\ \hline
8 & 0.0604955 & 0.0120903 \\ \hline

\end{tabular}
\end{center}

\vspace{0.3cm}
\begin{center}
\scalebox{.9}{\includegraphics{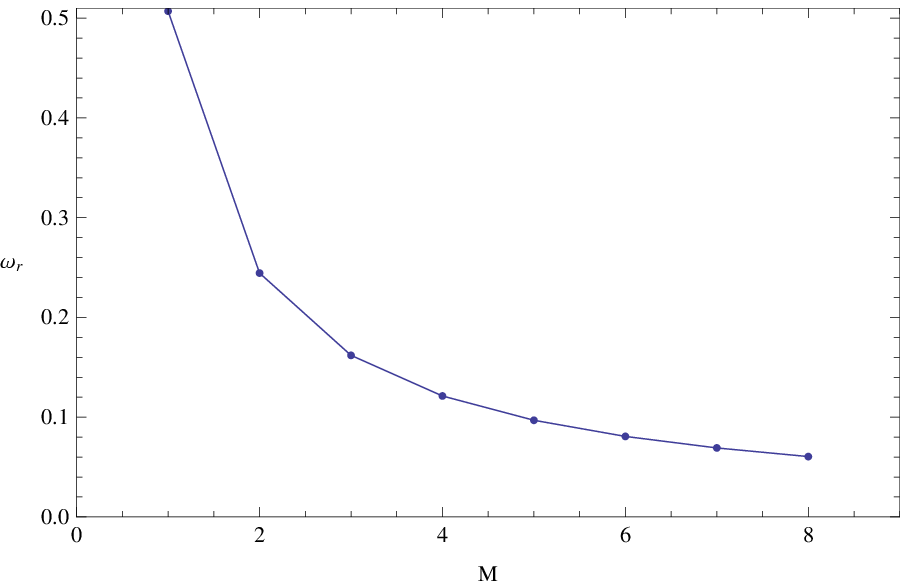}}

\vspace{0.3cm}

\end{center}

Figure 10. The behavior of Re $\omega$ with the mass  $M$ for $l=2$, and $q=0.5$.\\

\begin{center}
\scalebox{.9}{\includegraphics{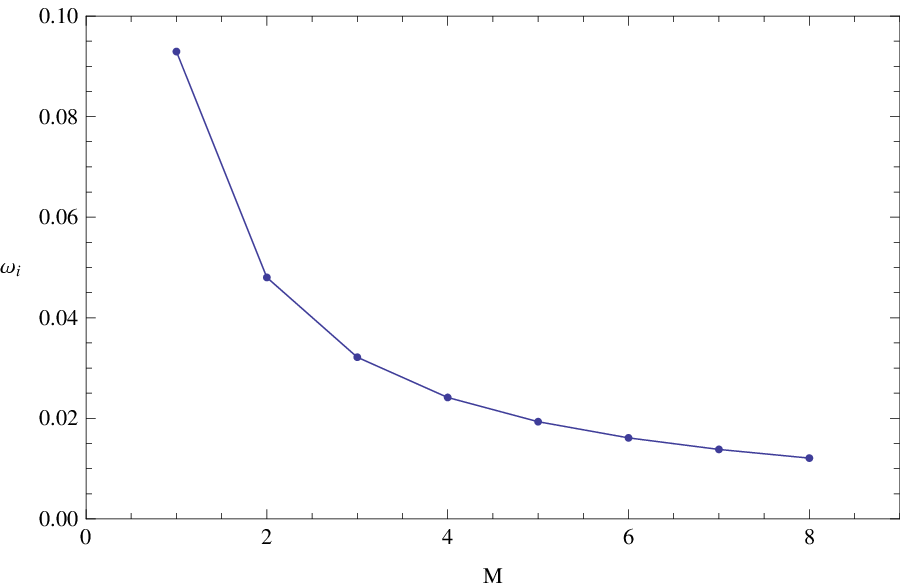}}

\vspace{0.3cm}

\end{center}

Figure 11. The behavior of Im $\omega$ with the mass  $M$ for $l=2$, and $q=0.5$.\\

When mass $M$ is increases, both $\omega_R$ and $\omega_I$ decreases.\\

Next, the QNM values are computed for the Bardeen black hole for various values of the spherical harmonic index  $l$ as shown in the following figures, Fig.12, Fig. 13, Fig. 14  and Fig.15. Note that the WKB work only for $ l > n$. Hence, we have chosen $l =2$ and computed $\omega$ of the fundamental mode with  $ n =0$ and the first over tone with $ n=1$.

\begin{center}
\begin{tabular}{|l|l|l|l|l|r} \hline \hline
 l & $\omega_R$ (n =0)  &  $\omega_I$( n =0) & $\omega_R$ ( n =1) & $ \omega_I$ ( n =1) \\ \hline

1 & 0.307491 & 0.0938013 &  **   &  **  \\ \hline
2 & 0.507037 & 0.0929337 & 0.490927     &   0.282895    \\ \hline
3 & 0.707780 & 0.0927166 &   0.695831    &   0.280290   \\ \hline
4 & 0.908912 & 0.0926315 &  0.899476  &   0.279202 \\ \hline
5 & 1.110220 & 0.0925589 & 1.102440   & 0.278647   \\ \hline
6 & 1.131162 & 0.0925651 & 1.305010   & 0.278327   \\ \hline
7 & 1.513080 & 0.0925500  &  1.507330   & 0.278125    \\ \hline
8 & 1.714570 & 0.0925399 & 1.709490 & 0.277990   \\   \hline
14 & 2.92388 & 0.0925168 & 2.920890    & 0.277678   \\ \hline
20 & 4.13340 & 0.0925107 & 4.131280    & 0.277596   \\ \hline
35 & 7.15741 & 0.0925067 &  7.156190   &  0.277541  \\ \hline
50 & 10.1815 & 0.0925057 & 10.18020    &   0.277528 \\ \hline

\end{tabular}
\end{center}

\vspace{0.3cm}
\begin{center}
\scalebox{.9}{\includegraphics{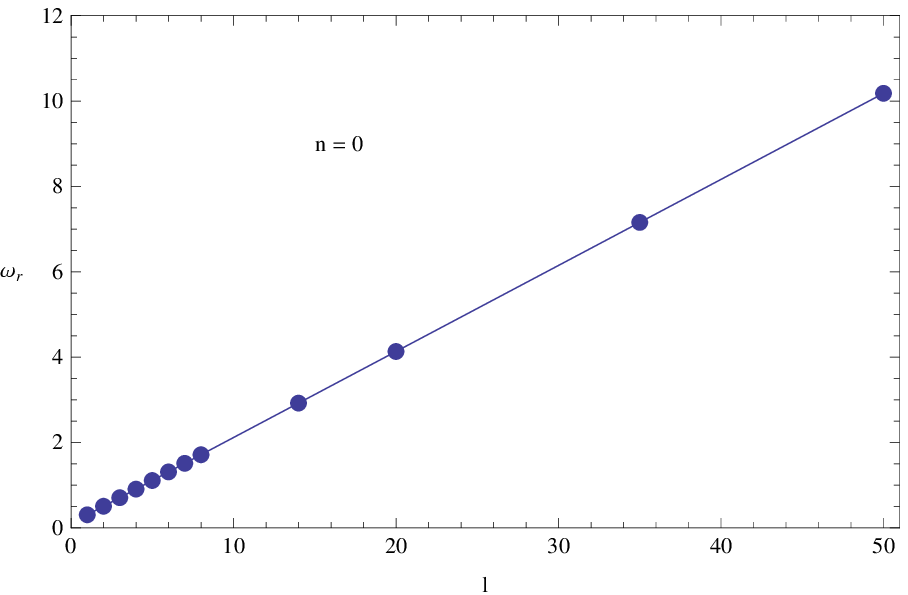}}

\vspace{0.3cm}

\end{center}

Figure 12. The behavior of Re $\omega$ with the   spherical harmonic index $l$ for $M=1$, and $q=0.5$. Here $n =0$\\

\begin{center}
\scalebox{.9}{\includegraphics{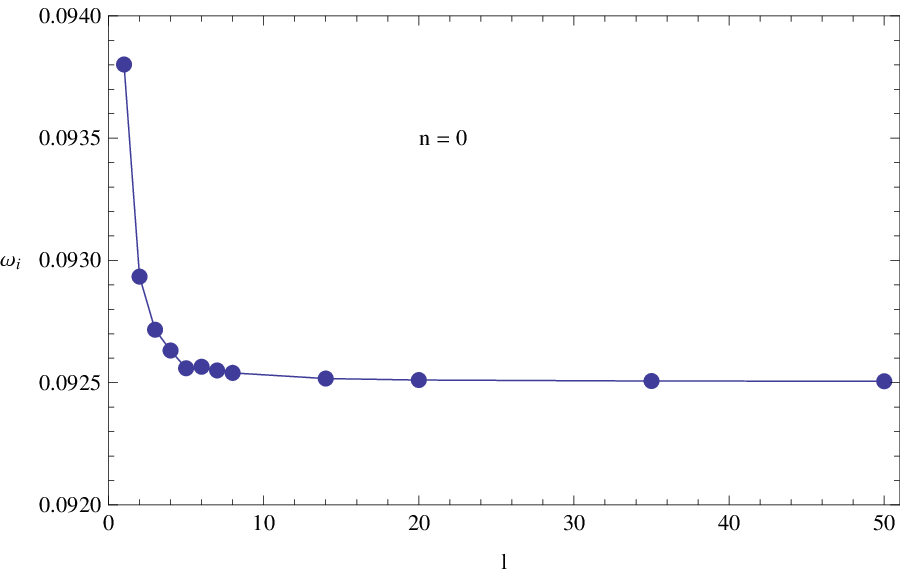}}

\vspace{0.3cm}

\end{center}

Figure 13. The behavior of Im $\omega$ with the spherical harmonic index  $l$ for $M=1$, and $q=0.5$. Here $n=0$.\\

\newpage

\vspace{0.3cm}
\begin{center}
\scalebox{.9}{\includegraphics{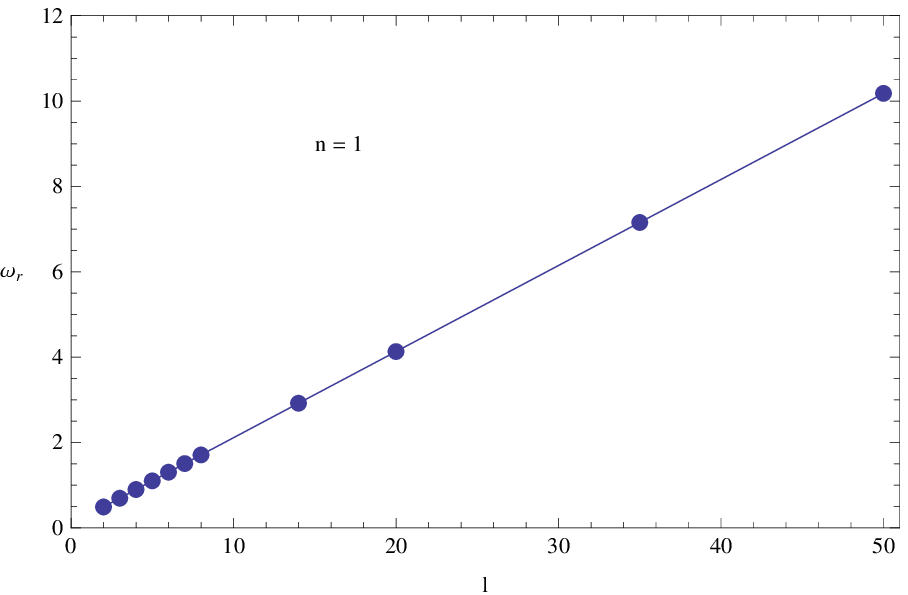}}

\vspace{0.3cm}

\end{center}

Figure 14. The behavior of Re $\omega$ with the   spherical harmonic  index $l$ for $M=1$, and $q=0.5$. here $ n =1$.\\

\begin{center}
\scalebox{.9}{\includegraphics{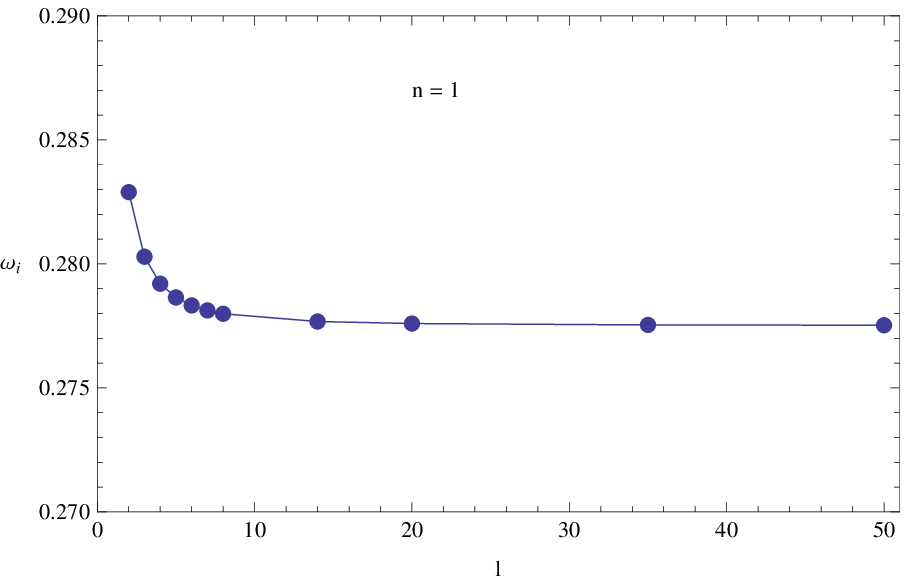}}

\vspace{0.3cm}

\end{center}

Figure 15. The behavior of Im $\omega$ with the spherical harmonic index  $l$ for $M=1$, and $q=0.5$. here $ n =1$.\\

For both $ n=0$ and $ n=1$, $\omega_R$ increases linearly with $l$. On the other hand, $\omega_I$ decreases and becomes stable for large $l$.

\section{ QNM's of the massless black holes in the eikonal limit from the null geodesics of the black hole}

\subsection{$\omega$ to the lowest order via null geodesics}

Cardoso et.al \cite{car} presented an important result to compute the QNM frequencies at the eikonal limit via the unstable null geodesics of the black hole for asymptotically flat black holes. This method was based on some earlier work done along these lines by Mashhon et.al\cite{mash1} \cite{mash2}. This approach has been applied to the Kerr black hole by Dolan\cite{dolan}, near extreme Kerr black hole by Hod \cite{hod1} and to the  black holes in anti-de-Sitter space by Morgan et.al \cite{morgan}.

First, let us give an introduction to the null geodesics of the Bardeen black hole. The geodesics of the Bardeen black hole were studied in detail by Zhou et.al \cite{zhou}. Hence referring to further details to that paper, we will only present the final equation of motion of the photons as,
\begin{equation}
\dot{r}^2 + V_{null}= E^2
\end{equation}
with,
\begin{equation}
V_{null} = \left(  \frac{L^2}{r^2}   \right) f(r) 
\end{equation}
Here, $L$ is the angular momentum of the photons. For $ r = r_h$, $ V_{null} = 0$ and for $ r \rightarrow \infty$, $ V_{null} \rightarrow 0$. In the Fig. 13, the  $V_{null}$ is given for various values of the magnetic charge $q$. The height is higher for higher charge $q$.

\begin{center}
\scalebox{.9}{\includegraphics{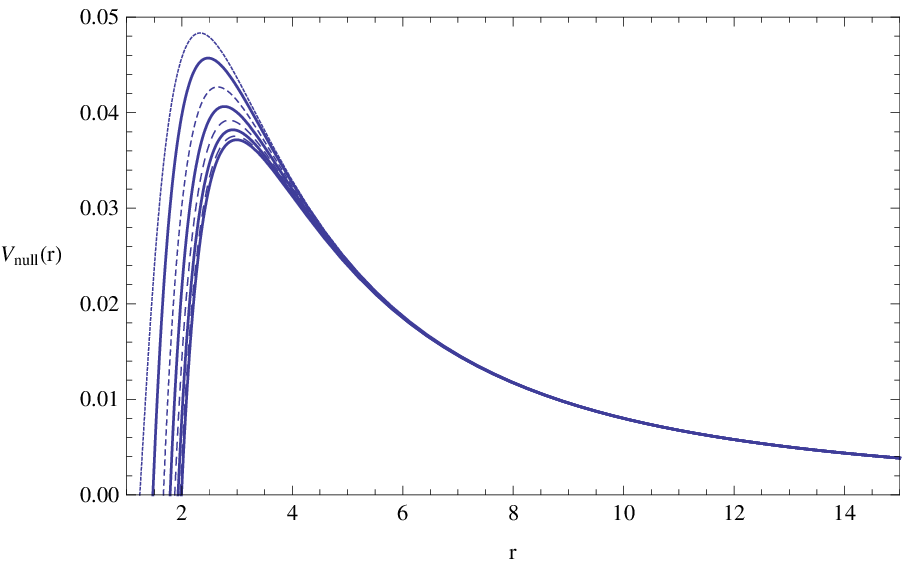}}\\
\vspace{0.1cm}
\end{center}
Figure 16. The graph shows the relation of $V_{null}$ with $r$ for various values of the magnetic  charge $q$. Here, $M =1$ and  $L =1$. When $q$ increases, $V_{null}$ increases.\\

Referring to the effective potential for the massless scalar field $V_{scalar}$ in eq.(17), one can see that in the eikonal limit ($ l \rightarrow \infty$),
\begin{equation}
V_{scalar}  \approx  \frac{ f(r) l}{r^2}
\end{equation}
Hence,  one can conclude that the maximum of $V_{scalar}$ occurs at $r = r_m$ given by,
\begin{equation}
2 f(r_m)  - r_m f'(r_m) =0
\end{equation}
Since the effective potential for the null geodesics, is given by $V_{null} = \frac{L^2 f}{r^2}$, the maximum of $V_{null}$  occurs at $V_{null}'=0$ leading to,
\begin{equation}
2 f(r_c)  - r_c f'(r_c) =0
\end{equation}
Hence the maximum of $V_{scalar}$ and the location of the maximum of the null geodesics coincides at $ r_m = r_c$.  Cardoso et.al.\cite{car} presented the QNM frequencies in the eikonal limit, as,
\begin{equation}
\omega_{QNM} = \Omega_c l - i ( n + \frac{1}{2} ) | \lambda|
\end{equation}
Here, $\Omega_c$ is the coordinate angular velocity given as,

\begin{equation}
\Omega_c =  \frac{\dot{\phi}}{\dot{t}}
\end{equation}
and,  $\lambda$ is the Lyapunov exponent which is interpreted as  the decay rate  of the unstable circular null geodesics. The derivation of the above results  is clearly given in Cardoso et.al\cite{car}. For the Bardeen black hole, $\Omega_c$ and $\lambda$ are given as,
\begin{equation}
\Omega_c = \frac{ \dot{\phi(r_c)}}{\dot{t}(r_c)}  = \sqrt{ \frac{ f(r_c)}{r_c^2 }} 
= \sqrt{ \frac{ ( r_c^2 + q^2)^{3/2}  - 2M r_c^2}{r_c^2(r_c^2 + q^2)^{3/2}}}
\end{equation}

\begin{equation}
\lambda = \sqrt{ \frac{ -V_{null}''(r_c)}{ 2 \dot{t}(r_c)^2}} 
= \sqrt{ \frac{-V_{null}''(r_c) r_c^2 f(r_c)}{2 L^2} }
\end{equation}
From the values of  $\Omega_c$ and $ \lambda$, one can extract the real and the imaginary part of $\omega$ from eq.(25) easily.  In the Fig. 17 and Fig. 18, $\Omega$ and $\lambda$ are given for the Bardeen black hole. One can conclude that the behavior is very similar when $\omega$ was computed using the WKB approach in section 4.

\begin{center}
\scalebox{.9}{\includegraphics{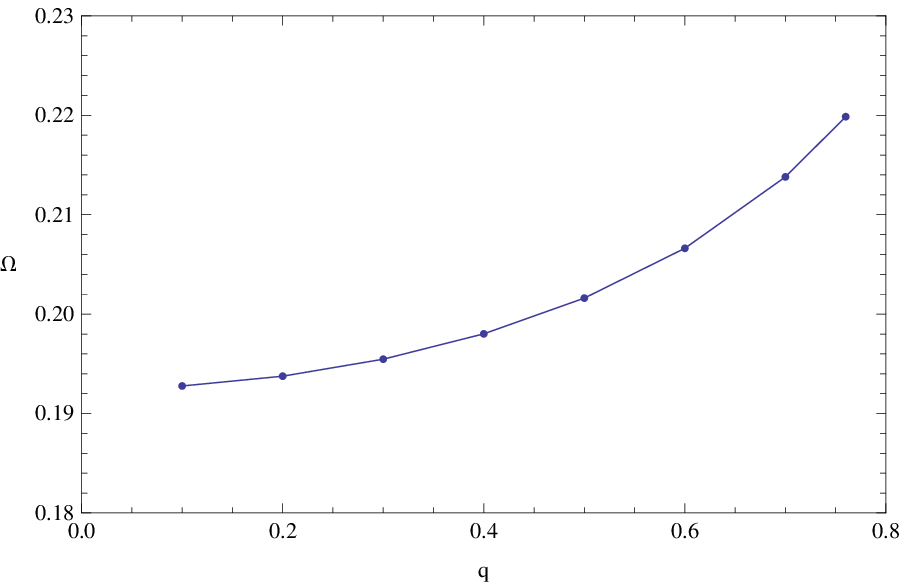}}\\
\vspace{0.2cm}
\end{center}
Figure 17. The graph shows $\Omega_c$ as a function of $q$. Here, $M = 1$ and $L=1$\\

\begin{center}
\scalebox{.9}{\includegraphics{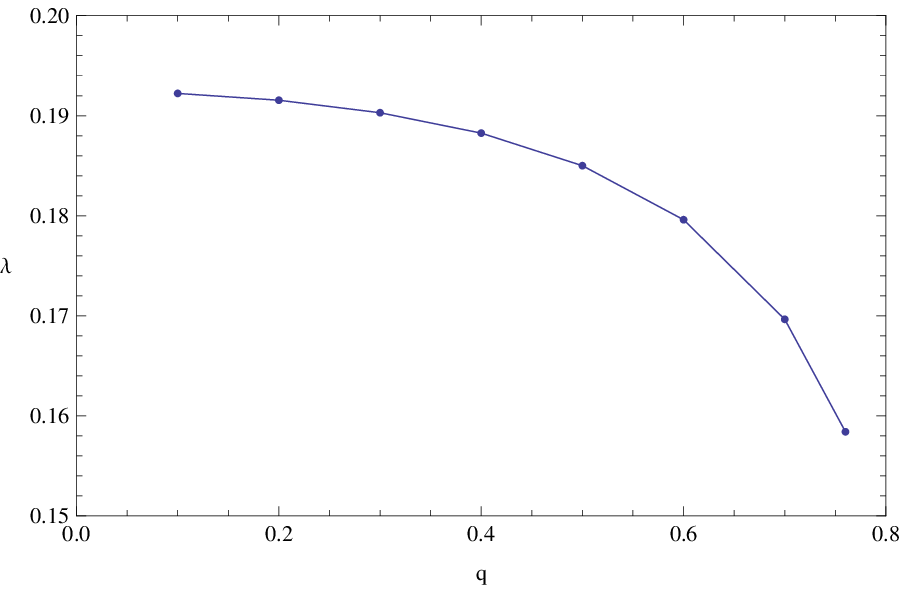}}\\
\vspace{0.2cm}
\end{center}
Figure 18. The graph shows the Lyapunov exponent $\lambda$ as a function of $q$.
Here, $M = 1$ and $ L =1$.\\

One can compare the $\omega$ in the eikonal limit for the Reissner-Nordstrom black hole with the Bardeen black hole. First let us  compute  the $r_c$ for the maximum of the effective potential for the  photons of the Reissner-Nordstrom black hole as,
\begin{equation}
r_c = \frac{ 1}{2} \left( 3 M + \sqrt{ 9 M^2 - 8 Q^2} \right)
\end{equation}
Hence, $\lambda_{RN}$ and $ \Omega_{RN}$ can be computed with the eq.(27) and eq.(28) with the corresponding metric function $f_{RN}(r)$. In the Fig.19 and Fig.20 $\lambda_{RN}$ and $\Omega_{RN}$ are plotted. Note that $\lambda_{RN}$ has a maximum at $ q = 0.7 M$. For the Reissner-Nordstrom black hole, the oscillation frequency, $\omega_R$, increases with the charge q.  The damping rate, $\omega_I$, increases and reaches a maximum at $ q = 0.7 M$ to  decrease rapidly to zero at $ M = q$. Ferrari and Mashhoon \cite{mash2} presented QNM's of the Reissner-Nordstrom black hole. In that paper, the QNM's were presented in the eikonal limit ( $ l \gg 1)$ and showed similar behavior. We like to note that the Lyapunov coefficients for the Reissner-Nordstrom black hole has been studied in detail in a recent paper by Pradhan \cite{pra}.

\newpage 

\begin{center}
\scalebox{.9}{\includegraphics{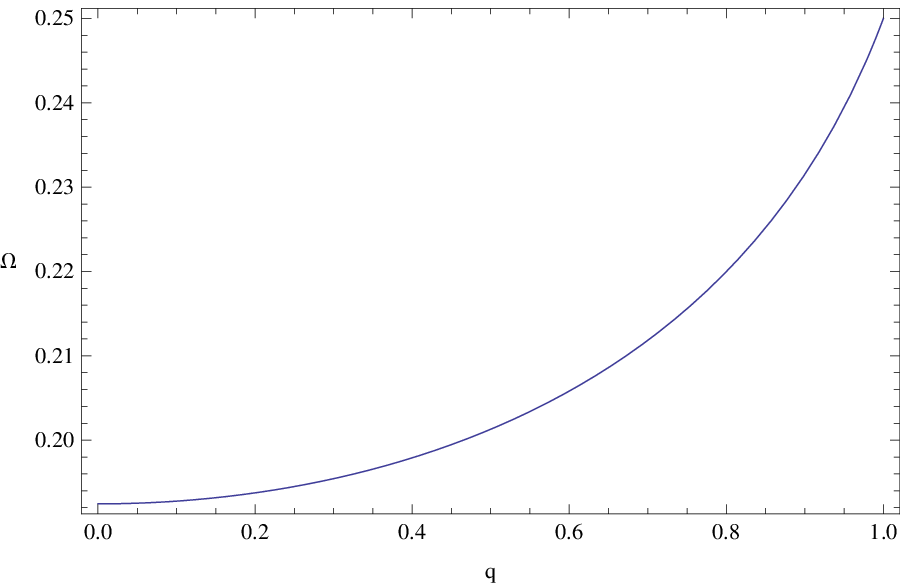}}\\
\vspace{0.2cm}
\end{center}
Figure 19. The graph shows $\Omega_c$ as a function of $q$ for the Reissner-Nordstrom black hole. Here, $M = 1$ and $L=1$ 

\begin{center}
\scalebox{.9}{\includegraphics{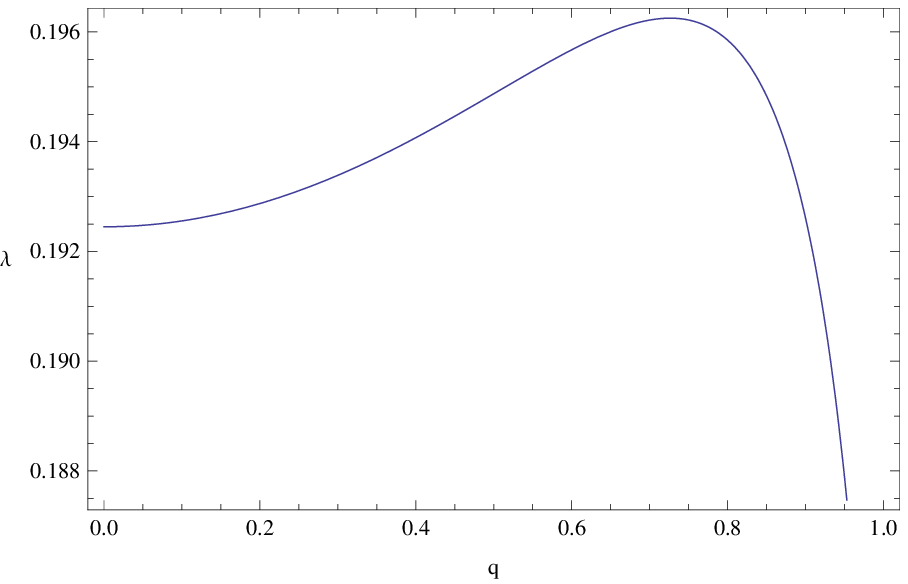}}\\
\vspace{0.2cm}
\end{center}
Figure 20. The graph shows the Lyapunov exponent $\lambda$ as a function of $q$ for the Reissner-Nordstrom black hole. Here, $M = 1$ and $ L =1$.\\

\subsection{On the expansion method to compute the QNM's at the eikonal limit}

Dolan and Ottewill \cite{sam2}  developed an expansion method to compute QNM's at the eikonal limit to higher orders.  Here, the basis for their approach is basically the same as given in section( 5.1) where  the null geodesics and the unstable circular orbits are the key to the computation. Other references where this method is applied are given in \cite{sam3} \cite{sam4} \cite{dolan}.

In this approach, the QNM's were expanded in inverse powers of $  L = l + 1/2$.
Let us first outline the formalism so that later we can apply it to the Bardeen black hole.  We shall start with the master equation for the scalar perturbation given in eq.(16),

\begin{equation}
\frac{d^2 \xi(r)}{dr_{*}^2} + \left( \omega^2  -  V_{scalar}(r_*) \right) \xi(r) =0
\end{equation}
where,
\begin{equation}
V_{scalar}(r) =  \frac{l(l+1) f(r)}{r^2}  + \frac{f(r) f'(r) }{r} 
\end{equation}
Now, redefine $\xi(r)$ as,
\begin{equation}
\xi(r) = exp^{\left( \int \frac{\alpha(r)}{f(r)} dr \right) }
\end{equation}
Here,
\begin{equation}
\alpha(r) = i b_c k_c(r) \omega
\end{equation}
The parameter $b$ is called the ``impact parameter'' given by $ L/E$ and $b_c$ is the value at $ r = r_c$ which is the radius of the unstable circular orbit. $L$ and $E$ of the null geodesics are conserved quantities of the orbits given by,
\begin{equation}
L = r^2 \phi ; \hspace{1 cm} E = f(r) t
\end{equation}
More information about the geodesics are given in  \cite{zhou} \cite{sam2} if one needs more details of the origin of these quantities.

Now, to define what $k_c(r)$ is, a new function $ k^2( r, b)$ is defined as,
\begin{equation}
k^2( r, b) = \frac{ 1} { b^2} - \frac{ f(r)} { r^2} 
\end{equation}
The origin of the function is at in the equation for null geodesics given in eq.(20) which also could be written as,

\begin{equation}
\left( \frac{ dr}{ d \phi} \right)^2 \frac{ 1}{r^2} = \frac{ 1} { b^2} - \frac{ f(r)} { r^2}
\end{equation}
In Dolan and Ottewill's  paper \cite{sam2}, the assumption is made that, there exists a critical impact paramter $b_c$, such that $ k^2(r, b_c)$ has degenerated roots. Since at $ r = r_c$, $dr/d \phi = 0$, this leads to,
\begin{equation}
k^2(r_c, b_c) =0; \hspace{1 cm}  \left.\frac{ \partial k^2(r, b_c)} { \partial r} \right|_{r = rc} =0
\end{equation}
If the assumption is made that the repeated root is a double root, a new function $k_c(r)$ is defined as,
\begin{equation}
k_c(r) = Sign( r - r_c) \sqrt{ k^2(r, b_c)} = ( r - r_c) K(r)
\end{equation}
Here, $K(r)$ becomes,
\begin{equation}
K(r) =  \sqrt{ \frac{ k^2(r, b_c)} { ( r - r_c)^2}}
\end{equation}
Hence, the function $k_c(r)$ is positive for $ r > r_c$ and negative for $ r < r_c$. If just  $ \sqrt{ k^2(r, b_c)}$ is considered to be $ k_c(r)$, then it would have lead to a function which is not differentiable at $ r = r_c$. On the other hand, the  definition given in eq.(38) leads to a smooth function at $ r = r_c$. To clarify these issues, we have plotted the graphs $k^2(r, b_c)$, $\sqrt{ k^2(r, b_c)}$, $ k_c(r)$ and $ d k_c(r)/dr$ in Fig.21   and Fig.22.

\begin{center}
\vspace{0.3 cm}

\scalebox{.9}{\includegraphics{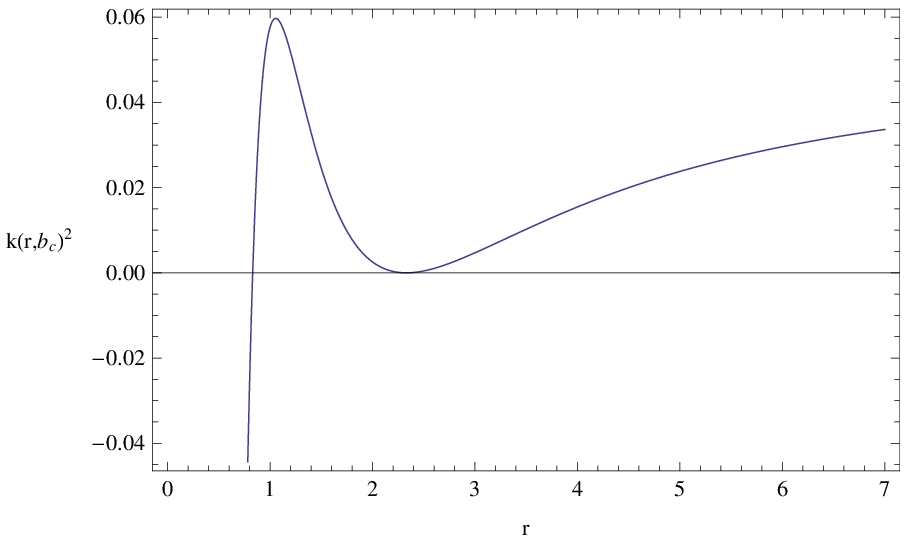}}

\vspace{0.3cm}
\end{center}

\begin{center}

\scalebox{.9}{\includegraphics{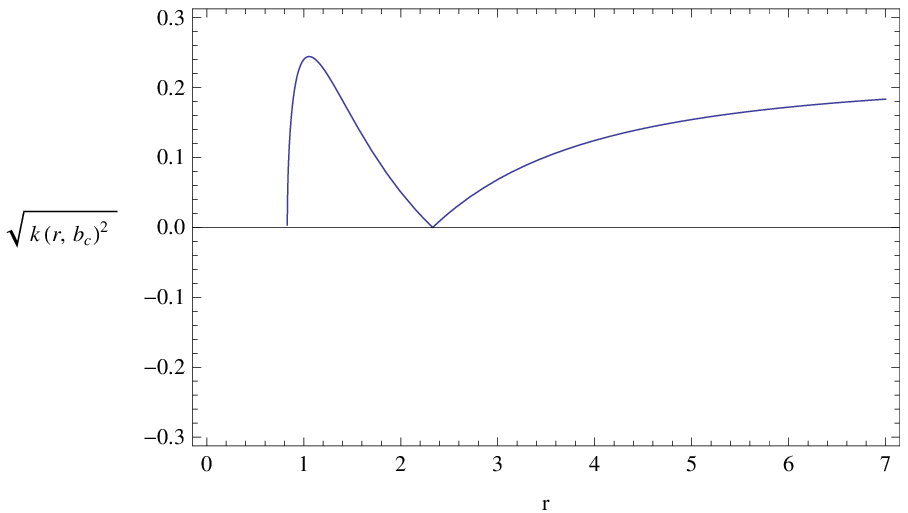}}

\vspace{0.3cm}
\end{center}

Figure 21. The behavior of $k(r, b_c)$  and $ \sqrt{ k(r, b_c)^2}$ with  $r$. Here $M=1$, $q=0.76$, $ r_c = 2.3299 $ and $ b_c = 4.5484 $.

\begin{center}
\vspace{0.3 cm}

\scalebox{.9}{\includegraphics{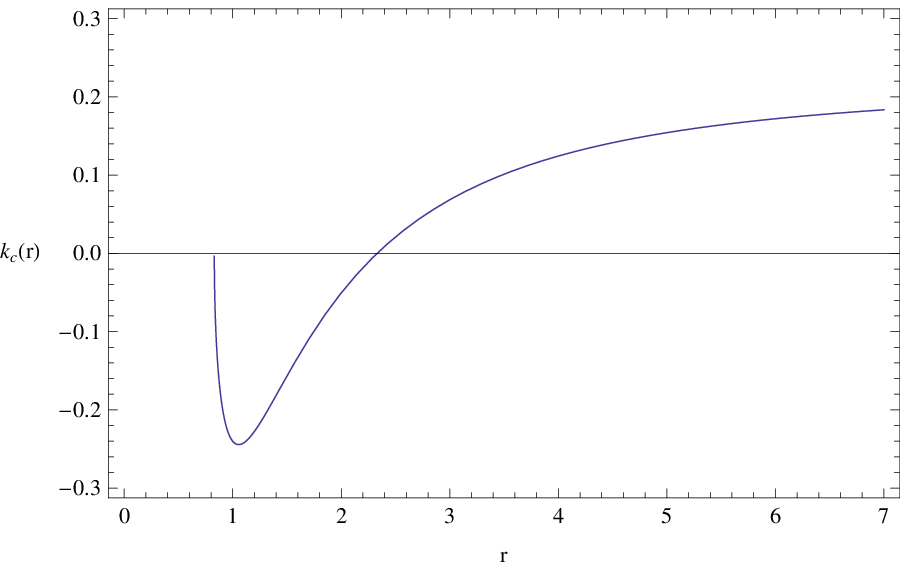}}

\vspace{0.3cm}
\end{center}

\newpage

\begin{center}

\scalebox{.9}{\includegraphics{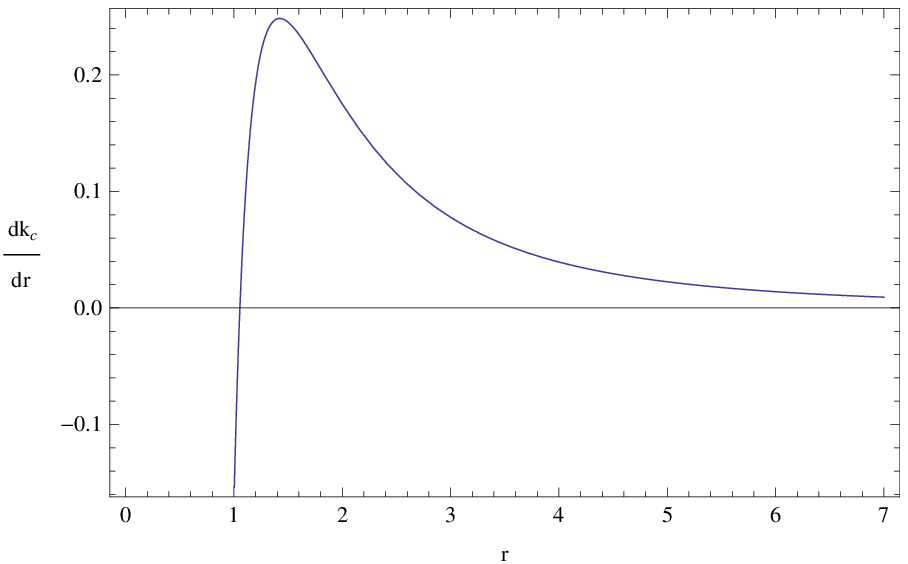}}

\vspace{0.3cm}
\end{center}

Figure 22. The behavior of $k_c(r)$  and $ \frac{dk_c(r)}{dr}$ with  $r$. Here $M=1$, $q=0.76$, $ r_c = 2.3299 $ and $ b_c = 4.5484 $.\\

Now that $k_c(r)$ is well established, on can substitute $\xi(r)$ into eq.(30) which simplifies to be,

$$
\frac{ d} { dr } \left( f(r) \frac{ dv(r)} { dr} \right)  + 
2 \alpha (r)\frac{df(r) } { dr} + $$
\begin{equation}
\left( \omega^2 + \alpha(r)^2 - V_{scalar}(r) + f(r) \alpha'(r) \right) \frac{ v(r) } { f(r)} =0
\end{equation}
Following the approach in \cite{sam2}, $\omega$ and $v(r)$ are expanded in inverse powers of $L$ as,

\begin{equation}
\omega _ { n =0} = L a_{-1} + a_{0} + L^{-1} a_1 + L^{-2} a_2 + .....
\end{equation}
and
\begin{equation}
v_{ n=0} = exp \left( S_0(r) + L^{-1} S_1(r)  + L^{-2} S_2(r) + .... \right)
\end{equation}
Note that in this paper, we are only interested in the fundamental modes corresponding to $ n =0$. The expansion could be done for higher modes with $ n >0$ as given in \cite{sam2}. Now, the $\omega$ and $v(r)$ are substituted to eq.(40) and similar powers of $ L$ are collected together as,
\begin{equation}
L^2:  \hspace{1 cm} a_1 = \sqrt{ \frac{f(r_c) }{ r_c^2}} = \frac{ 1}{ b_c}
\end{equation}
$$
L^1: \hspace{ 1 cm} \frac{ 2 a_{-1} a_0} { f(r) } - \frac{ 2 b_c^2 k_c(r)^2 a_{-1} a_{0} } { f(r)}+$$

\begin{equation}
i b_c a_{-1} k_c'(r) + 2 i b_c k_c(r) a_{-1} S_0'(r)=0
\end{equation}
\newpage

$$
L^2:  \left( \frac{ 1}{ 4 r^2} - \frac{ f'(r)}{r} \right) + \left( a_0^2 + 2 a_{-1} a_1 \right)  \frac{ (1 - b_c^2 k_c(r)^2 ) }{ f(r)}+ $$

$$ \left( a_0 S_0'(r) + a_{-1} S_1'(r) \right) ( 2 i b_c k_c(r) ) +$$

\begin{equation}
 i a_0 b_c k_c'(r) + f'(r) S_0'(r) + f(r) \left( S_0'(r)^2 + S_0''(r) \right)=0
\end{equation}
Here $'$ denotes differentiation with respect to $r$. The coefficients $a_{i}$ are found by imposing continuity condition at $ r = r_c$. Once $a_i$ are found, $S_i'(r)$ could be found. For example, from $ L^2$ terms, $ a_{-1}$ is evaluated at $ r = r_c$ as,
\begin{equation}
a_{-1} = \sqrt{ \frac{ f(r_c)}{r_c^2}}
\end{equation}
Note that this is same as $ \Omega_c$ computed in eq.(27).

From $L^2$ term, $a_0$ is evaluated at $ r = r_c$ as,
\begin{equation}
a_0 = \frac { - i b_c}{2}  f(r_c) k_c'(r_c)
\end{equation}
By substituting $ a_0$ and $a_{-1}$ back into eq.(44), one obtains $S_0'(r)$. One can differentiate $S_0'(r)$ to obtain $ S_0''(r)$ and substitute to eq.(45). Then, $a_1$ can be evaluated $ r = r_c$. This process can be continued to find all $a_i$ values and all $S_i'(r)$ functions. In this paper, we will only compute $ a_{-1}$, $ a_0$ and $a_1$ for the Bardeen black hole. We have chosen, $ M = 1$ and $ q = 0.76$ leading to $ r_c = 2.3299$ and $b_c = 4.5484$. By the expansion method, the following $a_i$ values are computed.

\begin{equation}
a_{-1} = 0.219858; \hspace{1 cm} a_0 = -0.0792063 i; \hspace{1 cm} 
a_1 =0.00499392
\end{equation}
Finally, $\omega_{n=0}$ in powers of $L$ is written as,

\begin{equation}
\omega_{n=0} = 0. 219858 L - 0.0792063 i  +  0.00499392 L^{-1}
\end{equation}

\section{Massive scalar perturbations}

In this section, we will address how the massive scalar field decay. For the
Schwarzschild black hole, it has been observed that the massive modes decay slower than the massless field \cite{lior} \cite{kono6}. Hence it is interesting to see if such  behavior is possible in the Bardeen black hole, leading to long lived modes.

Let us first present the  equation for a massive scalar field in curved space-time  as,
\begin{equation}
\bigtriangledown ^2 \Phi - m^2 \Phi =0
\end{equation}
Using the ansatz similar to eq.(15),
one obtain the  modified potential as,
\begin{equation}
V_{massive}(r) =  \frac{l(l+1) f(r)}{r^2}  + \frac{f(r) f'(r) }{r} + m^2 f(r)
\end{equation}
The effective potential $V_{massive}$ is plotted in Fig. 23  for  varying values of the mass $m$ of the scalar field.

\begin{center}
\vspace{0.3 cm}

\scalebox{.9}{\includegraphics{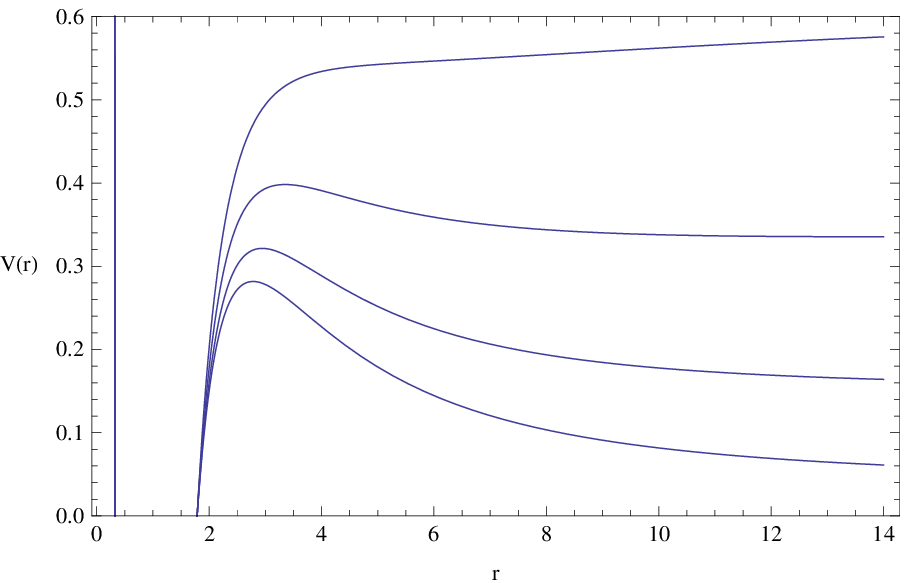}}

\vspace{0.3cm}
\end{center}

Figure 23. The behavior of $V_{massive}$ with the mass $m$ of the scalar field. Here $M=1$, $q=0.5$,  and $l=2$. When the mass decreases, the height of the effective potential also decreases.

The QNM's for the massive scalar field decay is computed using the WKB approximation discussed in Section 4. They are given in the following figures, Fig. 24 and Fig.25.

\begin{center}
\begin{tabular}{|l|l|l|l|l|r} \hline \hline
 m & $\omega_R$ (n =0)  &  $\omega_I$( n =0) & $\omega_R$ ( n =1) & $ \omega_I$ ( n =1) \\ \hline

0 & 0.551623 & 0.0796005 &  0.534227   &  **  \\ \hline
0.1 & 0.553910 & 0.0791170 & 0.536006     &   0.282895    \\ \hline
0.2 & 0.560821 & 0.0776240 &   0.541304    &   0.280290   \\ \hline
0.3 & 0.572499 & 0.0749856 &  0.550006  &   0.279202 \\ \hline
0.4 & 0.589201 & 0.0709443 & 0.561865   & 0.278647   \\ \hline
0.5 & 0.611305 & 0.0650541 & 0.576393   & 0.278327   \\ \hline
0.6 & 0.639305 & 0.0565300  &  0.590768   & 0.278125    \\ \hline
0.64 &  0.652259       &   0.0520869      &  0.586619 & 0.176782   \\   \hline
0.67 &   0.662598      &      0.0482116   &  0.542126    & 0.162060   \\ \hline
0.69 & 0.669602 & 0.0452134 & 0.377631    & 0.149858   \\ \hline

\end{tabular}
\end{center}

\vspace{0.3cm}
\begin{center}
\scalebox{.9}{\includegraphics{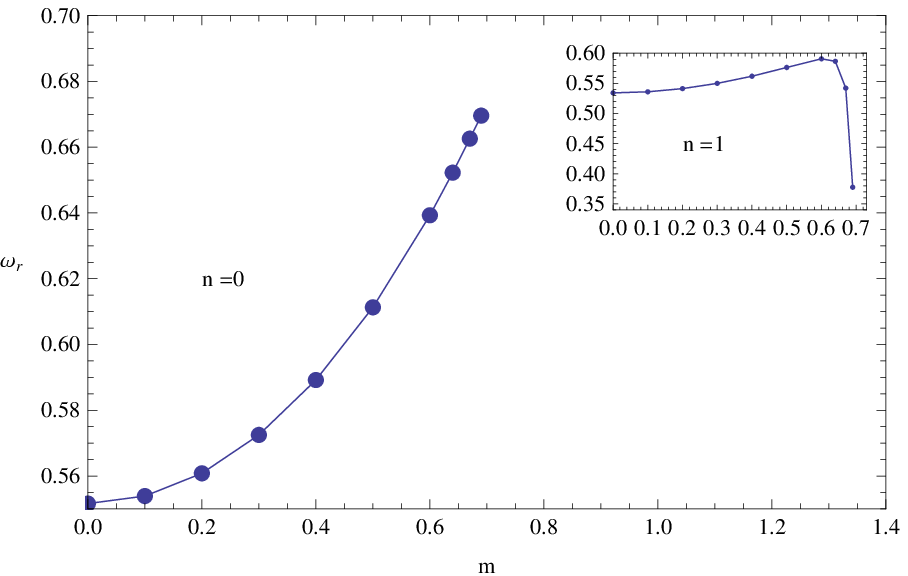}}

\vspace{0.3cm}
\end{center}

Figure 24. The behavior of Re $\omega$ with the  mass of the scalar field $m$ for $M=1$, $q=0.76$,  and $l=1$. Plots for $n =0$ and $n =1$ are given.\\

\begin{center}
\scalebox{.9}{\includegraphics{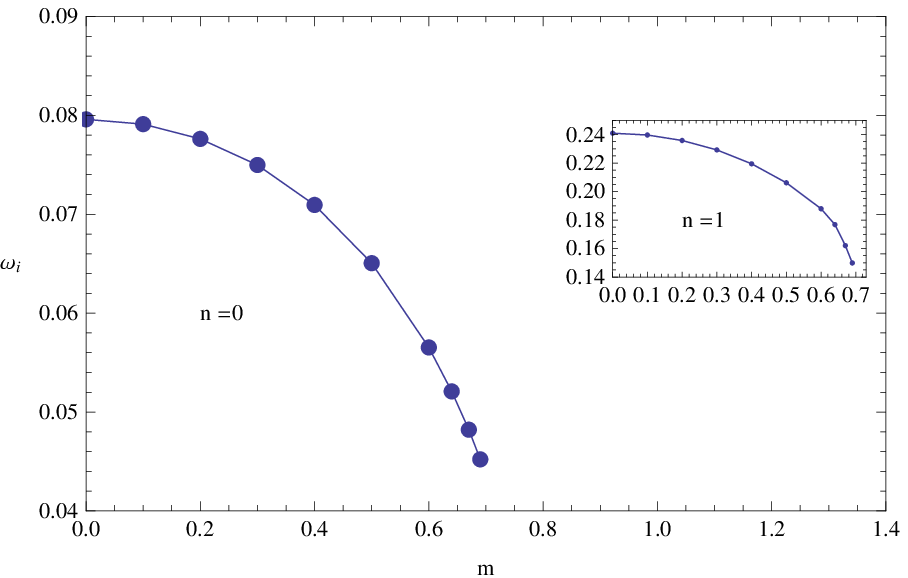}}

\vspace{0.3cm}

\end{center}

Figure 25. The behavior of Im $\omega$ with the mass of the scalar field $m$ for  $M=1$, $q = 0.76$, and $l=1$. Plots for $n =0$ and $n =1$ are given.\\

From Fig.24, the real part of the lowest QNM frequency increases with mass of the field. However, for the first overtone ( $ n=1$), the real part of QNM reaches a maximum and decreases rapidly with the mass of the field. Ohashi and Sakagami  did an analysis on the QNM's for the Reissner-Nordstrom black hole with the massive scalar field in \cite{ohashi}. The real part of the $\omega$ shows similar behavior as for the $\omega_R$  for  $ n=0$  for the Bardeen black hole. The imaginary part of the $\omega$ decreases with the mass for the Bardeen black hole for both values of $n$. This is similar to the behavior shown 
for $\omega_I$ of the Reissner-Nordstrom black hole \cite{ohashi}.

\section{Summary}

We have studied the scalar perturbations of the Bardeen black hole. Quasinormal mode spectrum of the massless scalar field is computed for various values of the charge $q$, mass $M$, and the spherical index $l$. The QNM spectrum is also computed for the Reissner-Nordstrom black hole by varying charge  along  with the Bardeen black hole and compared the behavior.

We have also applied the unstable null geodesics of the black hole to compute the QNM frequencies in the eikonal limit ( $ l \gg 1 $). Once again, a comparison is done with the QNM frequencies of the Reissner-Nordstrom black hole in the eikonal limit. We have also introduced the expansion method to compute $\omega$ in inverse powers of $L$ following the approach by \cite{sam2}.

Finally, the QNM frequency spectrum is computed for the massive scalar field. A discussion is also presented comparing the spectrum to the massive modes of the Reissner-Nordstrom black hole.

\section{Directions for further study}

There are several avenues to proceed from here to extend this work. For the Reissner-Nordstrom black hole, in the eikonal limit ($l \gg 1$), the effective potentials for scalar, Dirac and gravitational perturbations are approximately the same. This is clearly presented in the paper by Ferrari and Mashhoon \cite{mash2}. It would be interesting to do the analysis on gravitational perturbations and Dirac perturbations of the Bardeen black hole to see if similar behavior persists.

Highly damped asymptotic QNM's of the black hole is not studied in this paper. As mentioned in the introduction, that is a very active area of research on QNM's. It would be interesting to extend this work to study $\omega$ for large $n$ values. Motl and Neitzke \cite{motl}, did an analytical study to compute the asymptotic QNM's for the Reissner-Nordstrom black hole. It would be interesting to compare the asymptotic values of the Bardeen black hole with such values to understand how the nonlinear nature effects the physical properties of the black holes. 

It would be interesting to compute higher order  expansion of $\omega$ in the eikonal limit using the approach by \cite{sam2}.

An extension of the Bardeen blackhole with a non-linear electromagnetic source presented by Ay\'{o}n-Beato and Garc\'{i}a \cite{gar1} has been studied by Nomura  and Tamaki \cite{nom}. The metric of this black hole is given by,

\begin{equation}
ds^2 = -f(r) dt^2 + f(r)^{-1} dr^2 + r^2 ( d \theta^2 + sin^2(\theta) d \varphi^2)
\end{equation}
where,
\begin{equation}
f(r) = 1 - \frac{2M r^2}{ ( r^2 + q^2 ) ^{3/2} } + \frac{ q^2 r^2} { ( r^2 + q^2)^2}
\end{equation}
Nomura and Tamaki concluded that the QNM frequencies become pure imaginary when $ n \to \infty$. It would be interesting to see if such behavior persists in all non-linear sources such as the one presented in this paper.

With regard to the QNM's for the massive scalar field, it would be interesting to do an analytical study if possible, to find a bound on the mass $m$ when $\omega_I$ becomes zero.

\vspace{0.3 cm}

{\bf Acknowledgements:} SF like to thank R. A. Konoplya for providing the {\it Mathematica} file for the WKB approximation.

\end{document}